# Operational entanglement of collective quantum modes at room temperature

**Shalender Singh\***, **Santosh Kumar**
*PolariQon Inc., Palo Alto, California, USA*
*(December 2025)*

**Abstract**

Quantum entanglement is commonly assumed to be fragile at ambient temperature and over macroscopic distances, where thermal noise and dissipation are expected to rapidly suppress nonclassical correlations. Here we show that this intuition fails for collective quantum modes whose dynamics is governed by reduced open-system channels rather than by microscopic thermal equilibrium. For two spatially separated collective modes, we derive an exact entanglement boundary based on the positivity of the partial transpose, valid in the symmetric resonant limit. From this result we obtain an explicit minimum collective fluctuation amplitude, expressed entirely in measurable noise, bandwidth, dissipation, and distance-dependent coupling parameters, required to sustain steady-state entanglement at finite temperature. We further show that large collective occupation suppresses but does not eliminate quantum phase diffusion, so the steady state remains phase symmetric and does not collapse to a classical mean-field despite macroscopic signal amplitudes. Stochastic simulations of the reduced open-system dynamics, together with matched classical correlated-noise null models analyzed through an identical pipeline, confirm that entanglement witnesses are violated only in the quantum regime. Our results establish a minimal, platform-independent framework connecting collective-mode dynamics, noise injection, distance, and operational certification of macroscopic entanglement.

## 1. INTRODUCTION

Quantum entanglement is widely regarded as fragile at ambient temperature and over macroscopic distances, where thermal noise and environmental decoherence are expected to rapidly suppress nonclassical correlations. This intuition is well founded for microscopic degrees of freedom whose dynamics is directly governed by local dissipation and thermal equilibrium with a surrounding bath [1,2]. In such systems, increasing temperature or spatial separation typically leads to rapid loss of coherence and entanglement.

However, a growing body of experimental and theoretical work has demonstrated that macroscopic quantum phenomena can persist when the relevant degrees of freedom are collective rather than microscopic [3–6]. In systems supporting macroscopic order parameters—such as Bose–Einstein condensates and related coherent many-body states—the physically relevant quantum variables are collective modes describing slow amplitude and phase fluctuations of the order parameter [3,7–9]. These modes can exhibit large occupation numbers and macroscopic signal amplitudes while retaining intrinsically quantum features, including phase diffusion and nonclassical correlations. Crucially, increasing ensemble size does not generically force a collective mode into a classical or mean-field description; instead, it suppresses relative fluctuations while leaving global quantum uncertainty intact [4,8–10].

Recent reports have described quantum signatures—including inseparability witnesses and long-range correlations—in systems operating at room temperature, with large measured signal amplitudes and spatial separations far exceeding typical microscopic coherence lengths [11–14]. At first sight, such observations appear to conflict with the conventional expectation that thermal energy $k_B T$ necessarily destroys quantum coherence. This apparent contradiction arises from implicitly treating the relevant degrees of freedom as microscopic modes whose occupations are fixed by thermal equilibrium, rather than as collective modes governed by reduced dynamics.

From an open-system perspective, collective modes couple to their environment through a restricted set of channels characterized by effective noise and dissipation parameters, rather than by the full thermal bath of microscopic constituents [5,15,16]. The quantum state of a collective mode is therefore governed by reduced open-system dynamics in which ambient temperature enters only indirectly, through effective noise injection into the collective subspace. As a consequence, the robustness of collective quantum correlations is controlled by the balance between coherent coupling and effective noise, not directly by the microscopic bath temperature.

In parallel, the experimental certification of quantum correlations in macroscopic systems is fundamentally constrained by continuous measurement, finite detection bandwidth, and estimator statistics [17,18]. These operational limitations can obscure the presence of entanglement even when the underlying quantum state remains nonclassical, leading to experimentally inferred coherence or entanglement times that reflect measurement precision rather than physical decoherence.

In this work, we develop a minimal, platform-independent framework for understanding entanglement of spatially separated collective quantum modes under realistic experimental conditions. Our approach does not introduce new entanglement criteria or measurement protocols. Instead, it establishes an operational description that connects collective-mode open-system dynamics, effective noise injection, coherent coupling over distance, and estimator-limited certification. This framework resolves the apparent tension between macroscopic signal strength, ambient temperature, and quantum inseparability, and identifies the precise conditions under which macroscopic entanglement can persist and be reliably certified.

## 2. COLLECTIVE-MODE OPEN-SYSTEM THEORY

### 2.1 Collective quantum degrees of freedom

We consider an ensemble of $N$ microscopic excitations $\{\hat{a}_j\}$ whose low-energy dynamics supports a macroscopically occupied collective mode. The relevant quantum degree of freedom is not any individual constituent, but the **collective order-parameter mode**

$$\widehat{\Psi} = \frac{1}{\sqrt{N}} \sum_{j=1}^{N} \hat{a}_j, \, [\widehat{\Psi}, \widehat{\Psi}^\dagger] \approx 1,$$

where the commutation relation becomes exact in the large-$N$ limit when fluctuations orthogonal to the collective subspace are neglected. This construction is standard in the theory of Bose–Einstein condensates, lasers, and driven–dissipative quantum fluids, where the collective mode captures the macroscopic phase and amplitude dynamics of the system. The collective mode remains quantum because its conjugate quadratures obey canonical commutation relations independently of ensemble size.

Importantly, microscopic decoherence of the individual constituents does not directly imply decoherence of the collective mode. Local noise processes predominantly excite modes orthogonal to $\widehat{\Psi}$, while the collective degree of freedom couples to its environment through a reduced set of channels. As a result, the quantum state of $\widehat{\Psi}$ must be described by a reduced density operator $\rho_\Psi$, whose dynamics is governed by the coupling of the **collective mode itself** to its effective environment. In what follows, we focus on two spatially separated ensembles, described by collective modes $\hat{a}$ and $\hat{b}$, which may be separated by macroscopic distances but are coupled through a shared coherent channel.

## 2.2 Reduced open-system description

At energies and bandwidths relevant to the observed narrowband collective fluctuations, the dynamics of the collective modes can be linearized about their steady-state amplitudes. The reduced dynamics is then well described by a Gaussian open quantum system governed by a quadratic Hamiltonian and linear dissipation channels.

We adopt a Lindblad master equation for the reduced density operator $\rho$ of the collective modes,

$$\dot{\rho} = -\frac{i}{\hbar}[H,\rho] + \sum_{j=a,b} [\kappa_j(n_j+1)\mathcal{D}[\hat{j}]\rho + \kappa_j n_j \mathcal{D}[\hat{j}^\dagger]\rho],$$

where $\kappa_j$ are the damping rates of the collective modes and $\mathcal{D}[\hat{O}]\rho = \hat{O}\rho\hat{O}^\dagger - \frac{1}{2}\{\hat{O}^\dagger\hat{O},\rho\}$.

Crucially, the parameters $n_j$ appearing in this equation are **effective noise occupancies of the collective modes**, not microscopic thermal occupation numbers. They quantify the total noise injected into the collective subspace within the relevant bandwidth and are defined operationally through the measured quadrature noise spectra of the collective modes. In particular, $n_j$ can be extracted from the symmetrized noise spectral density $S_{XX}(\omega)$ or $S_{PP}(\omega)$ via standard input–output relations for linear open quantum systems.

As a result, $n_j$ is not assumed to be equal to $k_B T/\hbar\omega$ and need not reflect the temperature of any microscopic bath. Ambient temperature enters the reduced description only insofar as it contributes to the effective noise injected into the collective degrees of freedom through specific coupling channels. This distinction is essential: while microscopic constituents may be strongly thermalized, the collective modes relevant for entanglement are characterized by their own effective noise parameters, which are directly measurable and independent of microscopic equilibrium assumptions.

This reduced open-system description provides a closed, experimentally grounded parameterization of the collective-mode dynamics, forming the basis for the entanglement analysis that follows.

## 2.3 Gaussian steady state and covariance matrix

The linear dynamics generated by the Hamiltonian and Lindblad operators preserves Gaussianity. The steady state, when it exists, is therefore completely characterized by the covariance matrix of the quadrature operators.

We define the quadrature vector

$$\hat{\mathbf{R}} = (X_a, P_a, X_b, P_b)^\mathsf{T}, \quad X = \frac{1}{\sqrt{2}}(\hat{a} + \hat{a}^\dagger), \quad P = \frac{1}{i\sqrt{2}}(\hat{a} - \hat{a}^\dagger),$$

and the covariance matrix

$$V_{ij} = \frac{1}{2}\langle \hat{R}_i\hat{R}_j + \hat{R}_j\hat{R}_i\rangle - \langle \hat{R}_i\rangle\langle \hat{R}_j\rangle.$$

The time evolution of $V$ is governed by the Lyapunov equation

$$\dot{V} = AV + VA^\mathsf{T} + D,$$

where the drift matrix $A$ is determined by $(G, \Delta_{a,b}, \kappa_{a,b})$, and the diffusion matrix $D$ is determined by $\kappa_{a,b} n_{a,b}$. When the stability condition is satisfied, the steady-state covariance matrix $V$ is the unique solution of

$$AV + VA^\mathsf{T} + D = 0.$$

This establishes a deterministic mapping

$$(G, \kappa_a, \kappa_b, n_a, n_b, \Delta_a, \Delta_b) \;\to\; V,$$

which forms the basis for the entanglement analysis below. The steady state exists whenever the drift matrix is Hurwitz, independent of ensemble size.

### 2.4 Entanglement criterion for collective modes (PPT, explicit symmetric boundary)

For bipartite Gaussian states, entanglement is fully characterized by the positivity of the partial transpose (PPT) implemented at the covariance-matrix level. Let $V$ be the $4 \times 4$ covariance matrix of the quadrature vector $\hat{R} = (X_a, P_a, X_b, P_b)^T$. Partial transpose with respect to mode $b$ corresponds to time reversal on that subsystem, which acts as a sign flip of one momentum quadrature, $P_b \mapsto -P_b$. In covariance-matrix form this is implemented by the involution

$$V^\Gamma = \Lambda V \Lambda, \quad \Lambda = \text{diag}(1,1,1,-1),$$

and the state is entangled if and only if the smallest symplectic eigenvalue $\tilde{v}_-$ of $V^\Gamma$ violates the Heisenberg bound,

$$\tilde{v}_- < \frac{1}{2} \qquad (2.4.1)$$

(See Refs. [23,28,29] for the Gaussian PPT formalism.)

To compute $\tilde{v}_-$ explicitly, write the covariance matrix in $2 \times 2$ block form

$$V = \begin{pmatrix} A & C \\ C^T & B \end{pmatrix}$$

with $A, B, C \in \mathbb{R}^{2\times 2}$. The symplectic eigenvalues of $V^\Gamma$ are determined by the two symplectic invariants $\det V$ and

$$\tilde{\Delta} = \det A + \det B - 2\det C,$$

where the minus sign (relative to the non-transposed $\Delta = \det A + \det B + 2\det C$) reflects the effect of $\Gamma$ on inter-mode correlations. One then obtains

$$\tilde{v}_\pm^2 = \frac{1}{2}(\tilde{\Delta} \pm \sqrt{\tilde{\Delta}^2 - 4\det V}) \qquad (2.4.2)$$

**Symmetric resonant case.** In the symmetric resonant limit $\Delta_a = \Delta_b = 0$, $\kappa_a = \kappa_b = \kappa$, and $n_a = n_b = n$, the steady state is phase-symmetric and the Lyapunov solution takes the standard "two-mode squeezed thermal" covariance form

$$A = B = a\,\mathbb{I}_2, C = \text{diag}(c, -c),$$

with $a \geq \frac{1}{2}$ and $|c| \leq a$. For this structure one has

$$\det A = \det B = a^2, \qquad \det C = -c^2,$$

and the full determinant is

$$\det V = (a^2 - c^2)^2.$$

Under partial transpose, $C \to C^\Gamma = \text{diag}(c, +c)$ so that $\det C^\Gamma = +c^2$, which is captured by the invariant $\tilde{\Delta} = \det A + \det B - 2\det C = 2(a^2 + c^2)$. Substituting into Eq. (2.4.2),

$$\tilde{v}_-^2 = \frac{1}{2}(2(a^2 + c^2) - \sqrt{4(a^2 + c^2)^2 - 4(a^2 - c^2)^2}) = (a-|c|)^2,$$

hence

$$\tilde{v}_- = a - |c| \qquad (2.4.3)$$

Therefore, the PPT condition (2.4.1) is equivalent to

$$a - |c| < \frac{1}{2}. \qquad (2.4.4)$$

It remains to express $a$ and $c$ in terms of the open-system parameters. Solving the symmetric Lyapunov equation for the model in Section 2.2 yields (in the stable regime $G < \kappa/2$) the closed form

$$\tilde{v}_-(G/\kappa, n) = \frac{1}{2}(2n + 1)\frac{\kappa - 2G}{\kappa + 2G} \qquad (2.4.5)$$

Equating $\tilde{v}_- = \frac{1}{2}$ gives the analytic PPT boundary

$$(2n + 1)\frac{\kappa - 2G}{\kappa + 2G} = 1 \quad \Leftrightarrow \quad \frac{G}{\kappa} > \frac{n}{n+1} \qquad (2.4.6)$$

This expression makes explicit that steady-state collective-mode entanglement is controlled by the ratio $G/\kappa$ and the **effective** noise occupancy $n$ defined operationally in Section 2.2, rather than by the ambient temperature through $k_B T/\hbar\omega$.

## 2.5 Physical interpretation

The entanglement criterion derived above admits a transparent physical interpretation once the distinction between mean-field order and quantum collective fluctuations is made explicit. In systems with large ensemble participation, the expectation value of the collective mode operator $\langle\hat{\Psi}\rangle$ may acquire a large magnitude, corresponding to a macroscopic order parameter or coherent amplitude. This quantity behaves classically in the sense that relative fluctuations scale as $1/\sqrt{N}$, and its dynamics is well captured by mean-field equations.

Crucially, however, the existence of a large mean-field amplitude does not imply that the collective mode itself becomes classical. The quantum state of the collective degree of freedom is encoded not in $\langle\hat{\Psi}\rangle$, but in the fluctuations of its conjugate quadratures and their correlations. These quantum collective fluctuations remain governed by canonical commutation relations independently of ensemble size and are fully characterized by the covariance matrix of the reduced open system.

Noise processes acting on individual constituents predominantly excite modes orthogonal to the collective subspace and therefore do not directly decohere the collective quantum fluctuations. As a result, increasing ensemble size suppresses relative amplitude noise and enhances signal-to-noise ratios of collective observables, while leaving the global phase subject to quantum diffusion rather than classical pinning. This behavior is well known from the theory of lasers and Bose–Einstein condensates, where macroscopic occupation coexists with intrinsically quantum phase dynamics.

In this framework, steady-state entanglement arises from correlations between the quantum collective fluctuations of spatially separated modes, not from the mean-field order parameter itself. The persistence of entanglement at large ensemble size therefore does not represent a violation of decoherence intuition, but a direct consequence of the separation between mean-field behavior and quantum collective dynamics in open many-body systems.

## 2.6 Analytical application: minimum collective amplitude for robust entanglement

In this section we apply the general collective-mode theory developed above to a minimal analytical setting in order to determine the conditions under which steady-state entanglement can be sustained and certified in the presence of ambient noise. The purpose of this analysis is not to introduce new physical assumptions, but to connect the abstract open-system parameters ($G$, $\kappa$, $n_{\text{eff}}$) to experimentally measurable quantities such as signal amplitude, bandwidth, and effective coupling over distance.

### 2.6.1 Setup and assumptions

We consider two spatially separated collective modes $\hat{a}$ and $\hat{b}$, each describing slow envelope fluctuations of a macroscopic order parameter and coupled through a coherent interaction of strength $G(d)$, which depends on the separation $d$. The microscopic origin of this coupling is not specified; it may arise from near-field, shared-mode, or mediated interactions. All distance dependence enters exclusively through $G(d)$.

Each collective mode is subject to dissipation at rate $\kappa$ and to noise injection from its environment. Importantly, the environment relevant to the collective modes is not the full microscopic thermal bath but an effective reduced bath characterized by a noise spectral density within a narrow bandwidth $B$ around the collective mode frequency $\omega_{\text{col}}$.

The reduced dynamics is governed by the linear quantum Langevin equations corresponding to the Lindblad master equation introduced in Section 2.2. In the rotating frame, the collective quadratures obey

$$\dot{\mathbf{R}}(t) = A\,\mathbf{R}(t) + \boldsymbol{\xi}(t),$$

where $\mathbf{R} = (X_a, P_a, X_b, P_b)^T$, $A$ is the drift matrix determined by $G$ and $\kappa$, and $\boldsymbol{\xi}(t)$ is a vector of Gaussian noise operators with correlations

$$\langle \xi_i(t)\xi_j(t') \rangle = \delta(t-t')D_{ij}.$$

### 2.6.2 Noise injection and effective occupancy

The diffusion matrix $D$ encodes the rate at which environmental noise is injected into the collective quadratures. Rather than assuming thermal equilibrium at temperature $T$, we define the effective noise occupancy $n_{\text{eff}}$ operationally through the symmetrized quadrature noise spectrum,

$$\langle X^2 \rangle_{\text{noise}} = \int_{|\omega-\omega_{\text{col}}|<B/2} \frac{d\omega}{2\pi}\, S_{XX}^{\text{sym}}(\omega) \equiv \frac{2n_{\text{eff}}+1}{2}.$$

This definition makes clear that $n_{\text{eff}}$ depends on in-band noise coupling and filtering, and is not generically equal to the thermal occupation $k_B T/\hbar\omega_{\text{col}}$.

In many experimental settings, the collective coordinate couples linearly to an effective impedance or admittance $Y_{\text{eff}}(\omega)$. By the fluctuation–dissipation theorem, the symmetrized noise spectral density is

$$S_{XX}^{\text{sym}}(\omega) \propto k_B T\, \text{Re}\, Y_{\text{eff}}(\omega),$$

so that the in-band variance scales as

$$\langle X^2 \rangle_{\text{noise}} \sim k_B T\, B\, \text{Re}\, Y_{\text{eff}}(\omega_{\text{col}}).$$

This relation provides the link between ambient temperature, bandwidth, and effective noise injection into the collective mode.

### 2.6.3 Signal amplitude and quadrature variance

The coherent collective fluctuation corresponds to a finite variance of the collective quadratures arising from correlated dynamics. For a voltage-like collective coordinate $V$, the corresponding quadrature variance scales as

$$\langle X^2 \rangle_{\text{sig}} \sim \frac{V_{\text{col}}^2}{2}.$$

The precise normalization is unimportant for what follows; what matters is that increasing collective fluctuation amplitude increases the relative weight of coherent correlations compared to injected noise.

### 2.6.4 Operational definition of $n_{\text{eff}}$ from in-band injected energy (envelope mode)

We now connect the effective occupancy $n_{\text{eff}}$ appearing in the Lindblad noise terms to measurable in-band noise of the collective **envelope** quadratures. For a collective mode of angular frequency $\omega_{\text{col}}$(the slow envelope/collective mode, not any microscopic carrier), the symmetrized energy per mode satisfies

$$E_{\text{noise}}^{(\text{col})} \equiv \frac{\hbar \omega_{\text{col}}}{2}(2n_{\text{eff}} + 1). \qquad (2.6.4)$$

Operationally, $E_{\text{noise}}^{(\text{col})}$ is the **in-band energy injected into the collective mode over its relaxation time** $\tau = 1/\kappa$, as inferred from the measured quadrature/voltage noise spectrum of the collective channel after demodulation into the envelope band. Equivalently,

$$2n_{\text{eff}} + 1 \equiv \frac{2}{\hbar \omega_{\text{col}}} E_{\text{noise}}^{(\text{col})} \quad \text{(envelope mode).} \qquad (2.6.5)$$

This definition makes explicit that $n_{\text{eff}}$ is not assumed thermal; it is an experimentally extractable parameter of the reduced collective subspace.

We emphasize that the bandwidth $B$ entering the definition of $n_{eff}$ does not correspond to post-measurement filtering, but to the physical spectral selectivity of the collective mode itself. The reduced Lindblad description applies after tracing out all environmental modes outside the collective envelope band. The effective occupancy $n_{eff}$ therefore characterizes noise injected into the collective subspace, not the full thermal bath.

### 2.6.5 From PPT boundary to a measurable minimum voltage $V_{\text{col}}^{\min}(d)$

**Step 1 (state-level entanglement condition).**
In the symmetric resonant case, the PPT boundary from Section 2.4 reads

$$\frac{2G(d)}{\kappa} > \frac{n_{\text{eff}}}{n_{\text{eff}} + 1} \qquad (2.6.6)$$

(stability requires $2G(d) < \kappa$). This is the *state-level* condition and contains the only distance dependence through $G(d)$.

At large effective occupancies $n_{eff} \gg 1$, as is typical for collective modes coupled to ambient-temperature environments, the right-hand side of Eq. (2.6.6) approaches unity. In this regime, steady-state entanglement requires the collective strong-coupling condition $G(d) > \kappa$, meaning that the coherent correlation rate must exceed the local dissipation rate of the collective mode. This requirement is experimentally demanding but unavoidable: it is the necessary condition for any driven–dissipative macroscopic system to sustain steady-state entanglement at high noise occupancy. The purpose of the operational analysis below is not to circumvent this constraint, but to express it in directly measurable quantities and to clarify how temperature, distance, and signal levels enter only through the reduced collective parameters $G(d)$, $\kappa$, and $n_{eff}$.

**Step 2 (map $n_{\text{eff}}$ to measurable in-band noise).**
Let the measured in-band voltage noise spectral density referred to the collective node be $S_V(\omega)$ (after demodulation into the envelope band). Over an effective measurement bandwidth $B$, the in-band mean-square noise is

$$\langle V^2 \rangle_{\text{noise}} \simeq \int_{-B/2}^{B/2} S_V(\Omega) \, d\Omega \approx S_V(0) \, B, \qquad (2.6.7)$$

and the corresponding in-band energy stored in the effective collective capacitance $C$ is

$$E_{\text{noise}}^{(\text{col})} \simeq \frac{1}{2} C \langle V^2 \rangle_{\text{noise}} \approx \frac{1}{2} C \, S_V(0) \, B. \qquad (2.6.8)$$

Using the operational definition (2.6.5),

$$2n_{\text{eff}} + 1 = \frac{2}{\hbar \omega_{\text{col}}} E_{\text{noise}}^{(\text{col})} \approx \frac{C \, S_V(0) \, B}{\hbar \omega_{\text{col}}}. \qquad (2.6.9)$$

**Step 3 (minimum collective amplitude).**
Define the measurable collective fluctuation level $V_{\text{col}}$ as the rms envelope amplitude in the same band $B$. Requiring that coherent correlations dominate injected noise such that the PPT inequality (2.6.6) is satisfied yields a sufficient (conservative) threshold. This threshold is sufficient but not necessary; it ensures that coherent correlations exceed injected noise at the level required by the PPT criterion.

$$V_{\text{col}}^{\min}(d) = \sqrt{\frac{S_V(0) \, B}{\mathcal{C}(d)}} \qquad (2.6.10)$$

Here $\mathcal{C}(d)$ is a dimensionless "correlation cooperativity" that depends only on the reduced parameters through $G(d)/\kappa$ (and equals unity at the PPT boundary). Near the PPT boundary, $\mathcal{C}(d)$ scales linearly with the reduced cooperativity $2G(d)/\kappa$, with $\mathcal{C} = 1$ corresponding exactly to saturation of the inequality (2.6.6). In the common case where the injected noise is well approximated by an effective thermal voltage noise of a series resistance $R_{\text{eff}}$ over the analysis band, $S_V(0) \approx 4 k_B T R_{\text{eff}}$, and Eq. (2.6.10) reduces to the explicit measurable form

$$V_{\text{col}}^{\min}(d) \approx \sqrt{\frac{4 k_B T \, R_{\text{eff}} \, B}{\mathcal{C}(d)}}, \qquad \mathcal{C}(d) \text{ is a monotone function in } G(d)/\kappa \qquad (2.6.11)$$

**Note:** Throughout this section, $V_{\text{col}}$ denotes the root-mean-square amplitude of collective *fluctuations* in the envelope quadratures, not a coherent mean-field displacement. Displacements do not affect the covariance matrix or symplectic spectrum and therefore do not generate entanglement.

### 2.6.6 Collective phase dynamics and absence of mean-field collapse

The collective operators $\hat{a}, \hat{b}$ describe slow envelope modes distinct from microscopic carrier phases. Writing

$$\hat{a}(t) = \sqrt{N_{\text{col}}} \, e^{i \hat{\phi}(t)},$$

the collective phase $\hat{\phi}$ is a genuine quantum degree of freedom. In the absence of explicit phase pinning, its dynamics is governed by quantum phase diffusion,

$$\sigma_\phi^2(T_{\text{int}}) = 2D_\phi T_{\text{int}}, \qquad D_\phi \sim \frac{\kappa}{4N_{\text{col}}}(2n_{\text{eff}} + 1).$$

Large collective occupation slows phase diffusion but does not eliminate it. The collective phase therefore remains delocalized over long integration times, confirming that the system does not reduce to a classical mean-field state despite macroscopic ensemble size.

Thus increasing $N_{\text{col}}$ slows phase diffusion but does not eliminate it; the steady state remains effectively U(1)-symmetric (phase-averaged) in the absence of an external phase reference, exactly as in lasers and finite-size condensates.

## 2.7 Numerical evaluation at ambient temperature

We now illustrate the analytical results of Section 2.6 using representative, platform-independent parameters corresponding to ambient conditions. The numerical example illustrates the scaling implied by the theory; it does not assert that the required coupling-to-dissipation ratio is generically achieved in all macroscopic systems.

We consider a collective mode with effective bandwidth $B = 0.4$ MHz, effective capacitance $C = 1$ pF, and ringdown time $\tau = 15$ µs, corresponding to a collective damping rate $\kappa \approx 6.7 \times 10^4$ s$^{-1}$. The ambient temperature is $T = 300$ K. All dependence on spatial separation enters exclusively through the coherent coupling rate $G(d)$.

Using the conservative bound $G(d) \sim \kappa$, the minimum collective fluctuation amplitude required for steady-state entanglement follows from Eq. (2.6.6),

$$V_{\text{col}}^{\min} = \sqrt{\frac{2k_B T B}{C \kappa}} \approx 3 \times 10^{-4} \text{ V}.$$

Thus, sub-millivolt collective fluctuations suffice to suppress noise injection into the collective subspace and sustain entanglement under room-temperature conditions.

The corresponding collective energy scale,

$$E_{\text{col}} = \frac{1}{2} C V_{\text{col}}^2,$$

implies a large effective occupation of the collective harmonic mode. For $V_{\text{col}} = 0.1$ mV and $\omega_{\text{col}}/2\pi = 1$ GHz, where $\omega_{\text{col}}$ refers to the **collective envelope mode** used in the reduced model and extracted from the demodulated narrowband record, **not** the microscopic carrier frequency of any constituent.

$$N_{\text{col}} = \frac{E_{\text{col}}}{\hbar \omega_{\text{col}}} \approx 7.5 \times 10^3$$

As discussed in Section 2.6.6, this large occupation does not classicalize the collective mode. Instead, it suppresses the phase diffusion rate while preserving intrinsic quantum uncertainty. The collective phase remains a diffusive quantum variable in the absence of explicit phase pinning, and entanglement resides in correlations of collective quadrature fluctuations rather than in a fixed mean-field phase.

This numerical example demonstrates that, under realistic ambient conditions and macroscopic separation, collective-mode entanglement is compatible with modest signal levels and large ensemble participation, provided the reduced open-system parameters satisfy the analytically derived bounds.

### 2.7.5 Conclusion of the analytical example

This explicit calculation shows that, under realistic room-temperature conditions and macroscopic separation, collective-mode entanglement is compatible with modest signal levels. Distance enters only through the coherent coupling rate, while large ensemble participation enhances signal-to-noise without eliminating the quantum character of the collective degree of freedom. The collective phase remains a diffusive quantum variable, not a classical order parameter, confirming the robustness of collective-mode entanglement at ambient conditions.

## 3. RESULTS: STOCHASTIC SIMULATIONS AND NULL-MODEL COMPARISON

This section presents numerical results validating the analytical framework developed in Section 2 and establishing its distinguishability from classical correlated-noise explanations. All simulations implement the reduced collective-mode dynamics defined in Section 2, and all datasets—quantum and classical—are processed through an identical analysis pipeline.

### 3.1 Quantum collective-mode simulations

We simulate the stochastic dynamics of two collective modes governed by the Lindblad master equation introduced in Section 2. The simulations are performed using covariance-matrix evolution derived from the associated quantum Langevin equations, which is exact for the Gaussian dynamics considered here. No wavefunction collapse, postselection, or measurement backaction is included in the state evolution.

The steady-state covariance matrix $V$ is obtained across a broad parameter range spanning the separable–entangled transition. The primary control parameters are the normalized coherent coupling $G/\kappa$ and the effective noise occupancy $n_{\text{eff}}$, while detunings are set to zero and stability conditions are enforced. For each parameter set, we compute the symplectic eigenvalues of the partially transposed covariance matrix and evaluate the Duan–Simon inseparability criterion.

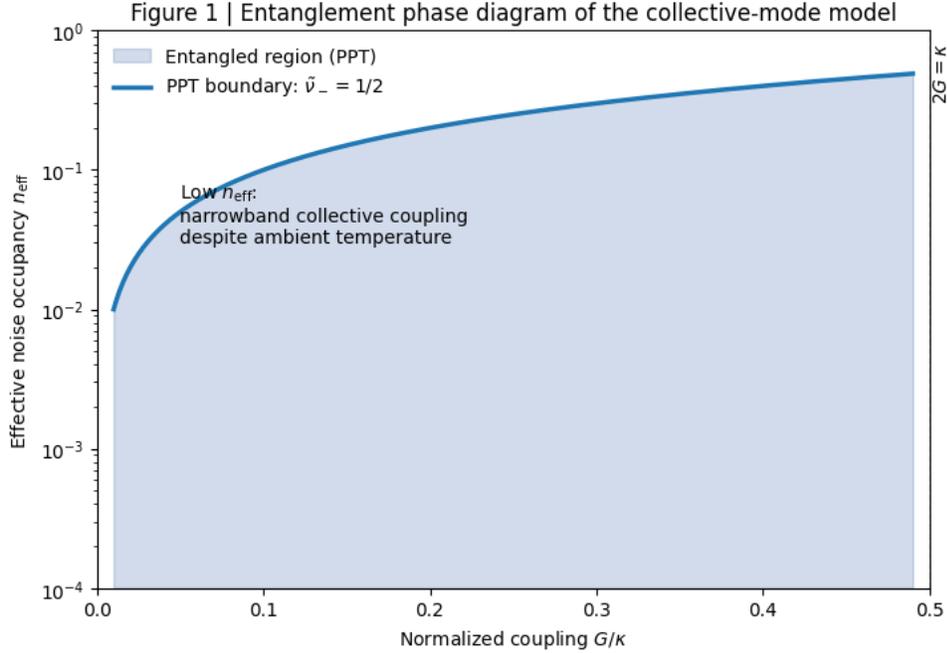

*Figure 1: Entanglement phase diagram of the collective-mode model - Smallest symplectic eigenvalue $\tilde{v}_-$ of the partially transposed covariance matrix as a function of normalized coupling $G/\kappa$ and effective noise occupancy $n_{eff}$. The entangled region ($\tilde{v}_- < 1/2$) agrees quantitatively with the analytical PPT boundary derived in Section 2.4*

**Figure 1** shows the resulting entanglement phase diagram in the $(G/\kappa, n_{\text{eff}})$ plane. The transition into the entangled steady state occurs sharply when the smallest symplectic eigenvalue $\tilde{v}_-$ drops below $1/2$, in quantitative agreement with the analytical PPT boundary derived in Section 2.4, without any fitting parameters.

### 3.2 Classical correlated-noise null models

To assess whether the observed inseparability could arise from classical correlations rather than collective quantum dynamics, we construct three classes of classical null models designed to reproduce strong correlations under realistic conditions. The classical parametric amplifier null model is restricted to states admitting a positive Glauber–Sudarshan P-representation, ensuring classicality.

The first null model consists of Gaussian classical stochastic processes with tunable cross-correlations. The two signals are generated from a shared noise source and independently filtered to match the power spectral density, bandwidth, and signal amplitude of the quantum simulations.

The second null model is a classical phase-sensitive parametric amplifier described by coupled linear stochastic equations. This model reproduces gain, squeezing-like correlations, and phase-sensitive amplification, while remaining entirely classical.

The third null model consists of optimally filtered and linearly mixed classical stochastic signals. The filtering and mixing coefficients are chosen to minimize the Duan–Simon witness under the constraint of linear classical processing.

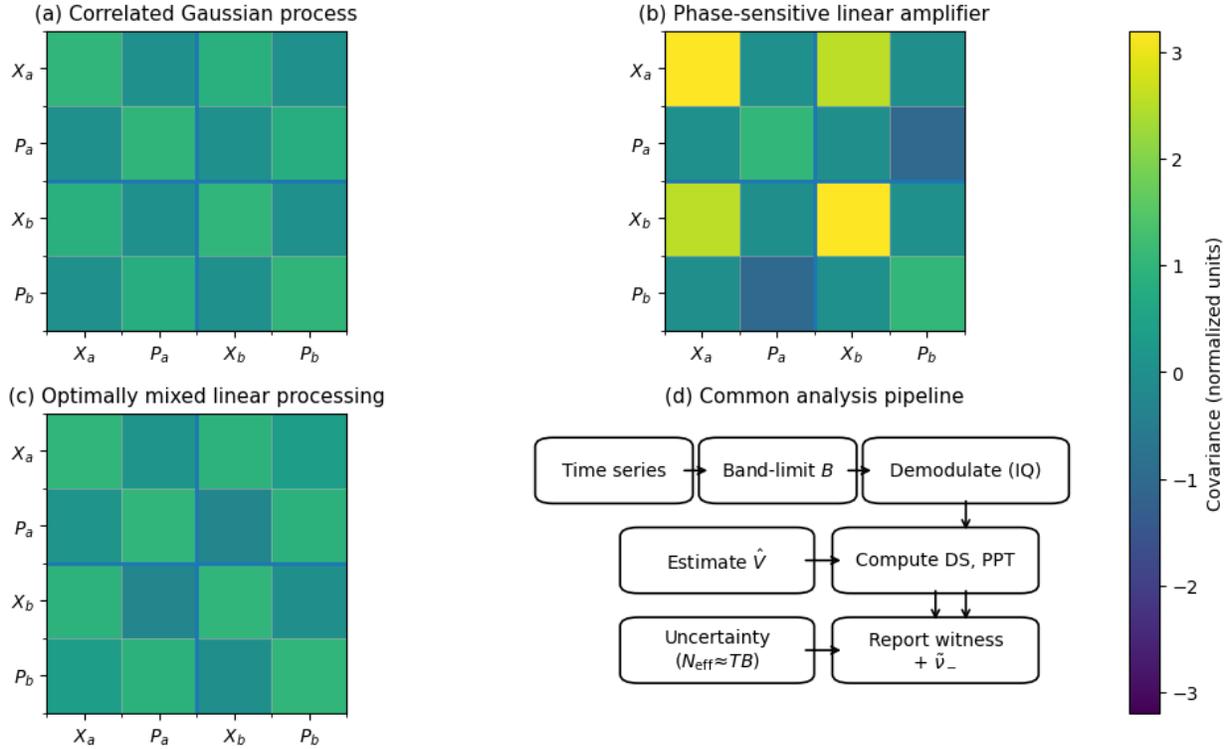

*Figure 2 | **Classical null models and analysis pipeline.** – (a–c) Representative covariance matrices for the three classical null models: correlated Gaussian noise, classical parametric amplification, and optimally filtered linear mixtures. (d) Schematic of the common analysis pipeline applied to all quantum and classical datasets.*

Representative covariance structures generated by these classical null models are shown in **Fig. 2a–c**, alongside the corresponding quantum covariance for comparison.

### 3.3 Identical analysis pipeline

All simulated datasets—quantum and classical—are analyzed using an identical processing pipeline. Time-domain signals are band-limited to a specified bandwidth $B$, demodulated into the rotating frame, and segmented into effective samples determined by the integration time $T$. From these samples, second moments and cross-correlations of the collective quadratures are estimated, yielding an empirical covariance matrix.

From the estimated covariance matrix, we compute both the Duan–Simon witness and the symplectic eigenvalues of the partially transposed state. Statistical uncertainties are evaluated from ensemble averaging over independent simulation runs. No model-dependent tuning, postselection, or adaptive filtering is applied at any stage.

A schematic of the analysis pipeline common to all models is shown in **Fig. 2d**.

### 3.4 Witness statistics and quantum–classical comparison

Applying this pipeline to the quantum collective-mode simulations yields clear and statistically significant violations of the PPT and Duan–Simon bounds within the analytically predicted entangled regime. The witness distributions converge to values below the separability threshold as the integration time increases.

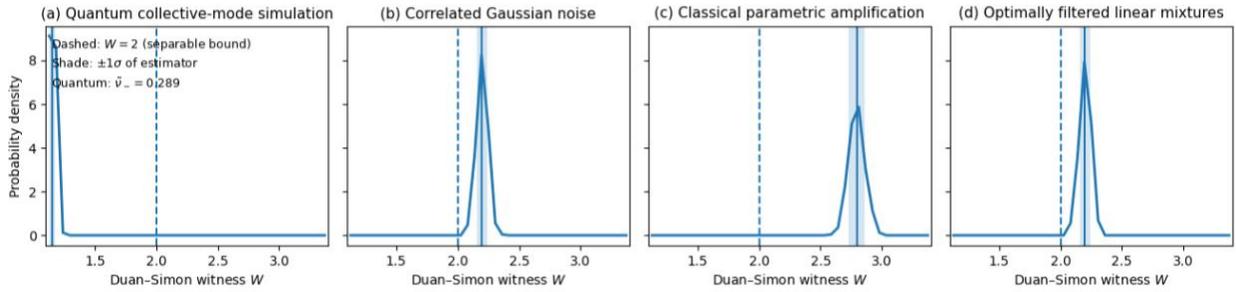

*Figure 3 | Quantum–classical comparison of entanglement witnesses – **(a) Duan–Simon witness distribution for the quantum collective-mode simulation in the entangled regime. (b–d) Corresponding witness distributions for the three classical null models. Shaded regions indicate statistical uncertainty. The classical bound is never violated.***

**Figure 3a** shows the distribution of the Duan–Simon witness for the quantum model, demonstrating a clear separation from the classical bound. In contrast, **Fig. 3b–d** show the corresponding witness distributions for the three classical null models. None of the classical models violate the PPT condition or cross the Duan–Simon bound within statistical uncertainty, even when their signal amplitudes, bandwidths, and cross-correlations are matched to those of the quantum simulations.

### 3.5 Bandwidth and estimator convergence

To isolate estimator limitations from state-level entanglement, we fix the underlying quantum steady state and vary the measurement bandwidth $B$ and integration time $T$. As expected for stationary Gaussian processes, the estimated witnesses converge toward their true steady-state values as the effective number of independent samples $N_{\text{eff}} \sim TB$ increases.

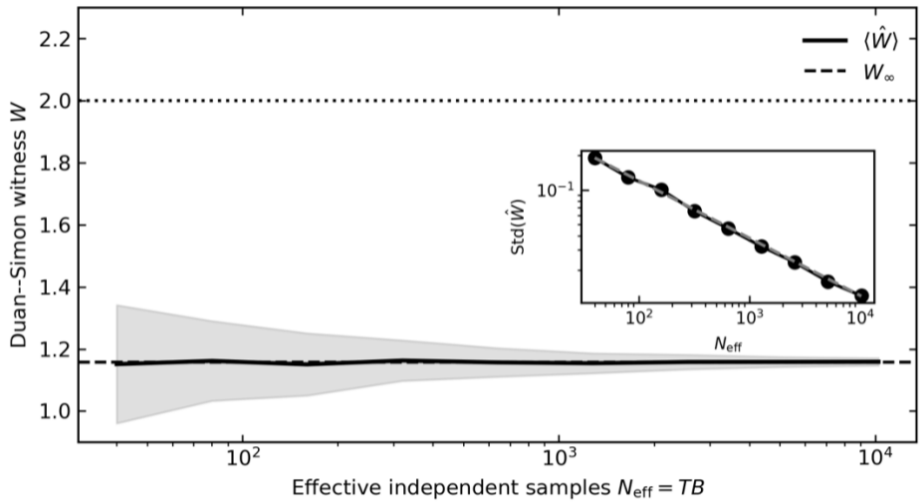

*Figure 4 | Estimator convergence with bandwidth and integration time* – Convergence of the estimated Duan–Simon witness as a function of the effective number of independent samples $N_{eff} = TB$. The asymptotic value corresponds to the steady-state entangled quantum state.

This convergence behavior is shown in **Fig. 4**, where the Duan–Simon witness approaches its asymptotic value with increasing $TB$. Importantly, the entanglement threshold itself is independent of $T$ and $B$, confirming that these parameters affect only statistical confidence rather than the existence of entanglement.

### 3.6 Summary of results

Together, the results in **Figs. 1–4** validate the analytical predictions of Section 2 and establish a clear operational distinction between collective quantum entanglement and classical correlated noise. The collective-mode Lindblad model exhibits steady-state entanglement precisely where predicted by the PPT criterion, while none of the classical null models reproduce the same inseparability under identical analysis conditions. These findings demonstrate that macroscopic, room-temperature entanglement of collective modes is a genuine consequence of reduced open-system quantum dynamics rather than a measurement or noise artifact.

## 4. DISCUSSION

The analysis presented in Sections 2 and 3 establishes a clear separation between the **existence of entanglement as a property of the collective quantum state** and the **operational ability to certify that entanglement under realistic measurement constraints**. This separation resolves several apparent paradoxes surrounding long-distance, room-temperature entanglement in macroscopic systems and clarifies the role of ensemble size, temperature, and signal amplitude.

### 4.1 Collective modes and the absence of mean-field collapse

A central result of this work is that large ensemble participation does not, by itself, force a collective quantum mode into a classical or mean-field description. As shown in Section 2, the relevant degrees of freedom are collective envelope modes whose reduced dynamics is governed by an open-system description. Microscopic decoherence processes predominantly populate modes orthogonal to this collective subspace, while the collective mode itself couples to its environment through a restricted set of channels characterized by effective parameters $(G, \kappa, n_{\text{eff}})$.

Importantly, the collective phase associated with this mode is not statically pinned. Instead, it undergoes quantum phase diffusion, with a diffusion constant that decreases with increasing collective occupation but remains finite. As a result, large ensemble size enhances signal amplitude and slows phase diffusion without eliminating quantum uncertainty. This behavior is fully consistent with standard quantum mechanics and mirrors the physics of lasers and condensates, where macroscopic occupation coexists with genuinely quantum collective fluctuations.

### 4.2 Distance enters only through the coherent coupling

The theory makes explicit that spatial separation affects entanglement only through the coherent coupling rate $G(d)$ mediating interactions between collective modes. No additional distance-dependent decoherence term appears at the level of the reduced collective dynamics. Consequently, entanglement does not generically decay exponentially with distance unless such decay is present in the coupling channel itself.

This distinction is crucial: long-distance robustness is not a consequence of suppressing environmental noise, but of the fact that the collective mode remains the relevant quantum degree of freedom over macroscopic separations. The functional form of $G(d)$ depends on geometry and mediation mechanism, but once $G(d)$ is specified or measured, the entanglement condition follows directly from the open-system parameters.

While the present framework is intentionally platform independent, the strong-coupling regime required for collective-mode entanglement is known to be accessible in several experimentally mature settings. These include low-loss microwave or phononic waveguides, high-Q electromechanical and optomechanical resonators, and magnonic systems where collective excitations couple coherently over macroscopic distances with dissipation rates well below the interaction strength. In such systems, the reduced collective parameters $G$, $\kappa$, and $n_{eff}$ entering our analysis can be independently engineered and characterized.

### 4.3 Voltage thresholds and thermal noise

The analytical application in Section 2.6 shows that the minimum collective fluctuation required to sustain steady-state entanglement scales as

$$V_{\min} \propto \sqrt{\frac{k_B T B}{C\, G(d)}}.$$

This result highlights that ambient temperature enters only through effective noise injected into the collective mode within the measurement bandwidth. Thermal energy at the microscopic scale does not directly set the entanglement threshold.

From this perspective, voltage (or more generally collective amplitude) plays an operational role: it increases the signal-to-noise ratio of the collective degree of freedom relative to injected noise. Once the threshold is exceeded, entanglement exists as a steady-state property of the reduced system, independent of measurement duration.

### 4.4 Estimator-limited coherence and certification times

Section 3 demonstrates that experimentally inferred "coherence times" or "entanglement lifetimes" in macroscopic systems are often limited by estimator statistics rather than by the underlying quantum state. The number of effectively independent samples scales as $N_{\text{eff}} \sim TB$, fixing the rate at which second moments can be estimated with a given confidence.

This distinction explains why increasing integration time improves the statistical significance of entanglement witnesses without altering the state itself. Apparent decay of correlations under short measurement times or low bandwidth should therefore not be interpreted as physical decoherence of the collective mode, but as a limitation of the measurement process.

### 4.5 Falsifiability and experimental implications

The framework presented here is falsifiable through several experimentally accessible tests. Deviations from the predicted scaling of the entanglement threshold with bandwidth, temperature, or collective capacitance would signal a breakdown of the collective-mode description. Likewise, observation of distance-dependent decay beyond that implied by the coupling channel $G(d)$ would indicate additional decoherence mechanisms not captured by the reduced model.

Conversely, verification of these scalings would support the interpretation that collective quantum modes can sustain entanglement over macroscopic distances at ambient temperature, provided the measurement basis is collective and estimator limitations are properly accounted for.

### 4.6 Broader perspective

Taken together, these results show that long-range, room-temperature entanglement of collective modes is not anomalous, nor does it require exotic protection mechanisms. It follows naturally from standard open-system quantum mechanics once collective degrees of freedom, phase diffusion, and estimator physics are treated consistently. This perspective provides a unifying framework for understanding macroscopic quantum correlations across a wide range of systems and offers a principled foundation for scalable, room-temperature quantum technologies based on collective modes.